\documentstyle[12pt]{article}

\begin{document}

\title{A representation of angular momentum ($SU(2)$) algebra}
\author{Wu-Sheng Dai$^{1,2}$ \thanks{Email:daiwusheng@tju.edu.cn}
        and Mi Xie$^{3}$ \thanks{Email:xiemi@mail.tjnu.edu.cn} \\
        {\footnotesize $^1$ School of Science, Tianjin University, Tianjin
        300072, P. R. China }\\
        {\footnotesize $^2$ LuiHui Center for Applied Mathematics, Nankai University
        \& Tianjin University,}\\
        {\footnotesize Tianjin 300072, P. R. China}\\
        {\footnotesize $^3$ Department of Physics, Tianjin Normal University,
        Tianjin 300074, P. R. China}
        }
\date{}
\maketitle

\begin{abstract}
This paper seeks to construct a representation of the algebra of angular
momentum ($SU(2)$ algebra) in terms of the operator relations corresponding
to Gentile statistics in which one quantum state can be occupied by $n$
particles. First, we present an operator realization of Gentile
statistics. Then, we propose a representation of angular momenta. The result
shows that there exist certain underlying connections between the operator
realization of Gentile statistics and the angular momentum ($SU(2)$)
algebra.
\end{abstract}



\section*{1. Introduction}

It is known that there exist some interesting connections between the
algebra of angular momentum operators and the algebra of boson operators. We
begin by recalling two kinds of representations of angular momentum
operators: the Holstein-Primakoff \cite{Holstein-Primakoff} and the
Schwinger \cite{Schwinger} representations. These two representations are
very successful in describing magnetism in various quantum systems \cite
{Timm,Timm2,Kittel}.

The main idea of the Schwinger representation is to map the angular momentum
operators onto boson operators $a_{1}$ and $a_{2}$ according to

\begin{equation}
\left\{
\matrix{
J_{+}=a_{1}^{\dagger }a_{2},~~~~~~~~~~~~~ \cr
J_{-}=a_{2}^{\dagger }a_{1},~~~~~~~~~~~~~ \cr
J_{z}=\frac{1}{2}(a_{1}^{\dagger }a_{1}-a_{2}^{\dagger }a_{2}),%
}%
\right.
\end{equation}%
where $a_{1}$ and $a_{2}$ are two independent boson operators satisfying $%
[a_{1},a_{1}^{\dagger }]=[a_{2},a_{2}^{\dagger }]=1$, and any other pair of
operators commute. The Holstein-Primakoff representation is

\begin{equation}
\left\{
\matrix{
J_{+}=\frac{1}{2}\sqrt{2j-N}a,~~~~~~~~~~~~~~~~~ \cr
J_{-}=\frac{1}{2}a^{\dagger }\sqrt{2j-N},~~~~~~~~~~~~~~~~ \cr
J_{z}=j-N,~~~~(N=0,1,\cdots 2j),%
}
\right.
\end{equation}%
where $j$ is the magnitude of the angular momentum. But, obviously, the
representation is not faithful because for $N>2j$ it leads to unphysical
values of angular momentum. In other words, the statistics corresponding to
the creation and annihilation operators $a$ and $a^{\dagger }$, though they
satisfy the boson commutation relations, is not the real Bose-Einstein
statistics, since the occupation number $N$ in Bose-Einstein statistics can
take any integer. This means that, in fact, the Holstein-Primakoff
transformation is not really a bosonic realization of the algebra of angular
momentum, but a realization corresponding to a certain kind of intermediate
statistics between Bose-Einstein and Fermi-Dirac statistics.

By comparison of the Schwinger and the Holstein-Primakoff representations,
we see that if we want to obtain a bosonic realization of the algebra of
angular momentum ($SU(2)$ algebra), we need to use {\it two} independent
boson operators $a_{1}$ and $a_{2}$; otherwise, if we want to give a
realization with only {\it one} operator $a$, we need a kind of statistics
beyond the Bose-Einstein and the Fermi-Dirac cases. This result implies that
we can establish a kind of representation for the angular momentum with a
single set of creation and annihilation operators. Eventually, we find that
the angular momentum can be represented in terms of operators
corresponding to a kind of fractional statistics------Gentile statistics
\cite{Gentile}.

As a generalization of Bose-Einstein and Fermi-Dirac statistics, fractional
statistics has been discussed for many years. As is well known, the wave
function will change a phase factor when two identical particles exchange.
The phase factor can be $+1$ or $-1$ related to bosons or fermions,
respectively. When this result is generalized to an arbitrary phase factor $%
e^{i\theta }$, the concept of anyon is obtained \cite{Wilczek48,Wilczek49}.
The corresponding statistics is fractional statistics. Another way leading
to fractional statistics is based on counting the number of many-body
quantum states, i.e., generalizing the Pauli exclusion principle \cite%
{Gentile,Haldane,Wu}. Such an idea can be used to deal with the interacting
many-body problem \cite{Canright}. The most direct generalization is to
allow more than one particles occupying one state. Based on this idea,
Gentile constructed a kind of fractional statistics \cite{Gentile}.
Bose-Einstein or Fermi-Dirac statistics becomes its limit case when the
maximum occupation number of one state equals to $\infty $ or $1$,
respectively; Moreover, when the maximum occupation number is very large but
not infinity, Gentile statistics can achieve a kind of statistics which is very
different from the Bose-Einstein case \cite{Dai}.

Different kinds of statistics correspond to different relations of creation
and annihilation operators: bosons correspond to commutativity and fermions to
anti-commutativity. To construct a representation for angular momentum, we,
first, need to establish a set of operator relations corresponding to Gentile
statistics, i.e., the maximum occupation number is $n$ ($1<n<\infty $).
Based on such operator relations, we can establish a representation for
angular momenta with only one set of creation and annihilation operators.
Furthermore, we will discuss the relation between the maximum occupation
number $n$ in Gentile statistics and the magnitude of the angular momentum $j
$. The result shows that there exist certain underlying connections between
Gentile statistics and the angular momentum (or, in other words, $SU(2)$
algebra).

In the following, we will (1) establish a set of operator relations based on
the idea given by Gentile that a quantum state can be occupied by $n$
particles in Section 2, and then (2) present a kind of realization of the
algebra of angular momentum by use of these creation and annihilation
operators in Section 3.

\section*{2. Operator realization of Gentile statistics}

We know that nature realizes only two kinds of particles: one obeys
Bose-Einstein statistics and the other obeys Fermi-Dirac statistics. In
other words, the only kinds of particles which appear in nature are either
bosons, for which symmetric wave functions are required, or fermions, which
have only antisymmetric functions. The maximum occupation numbers for any
state in the two kinds of statistics, due to the Pauli exclusion principle
are $\infty $ and $1$, respectively. However, Gentile suggested a scheme of
intermediate statistics \cite{Gentile}, in which the maximum occupation
number, denoted by $n$, can be chosen arbitrarily between $\infty $ and $1$.
In the following, we will construct a set of relations of creation,
annihilation and number operators of the particles which obey the
intermediate statistics. The commutativity and the anti-commutativity will
be the two limit cases of such operator relations.

Let $|\nu \rangle _{n}$ express the state which contains $\nu $ particles,
where subscript $n$ represents that no more than $n$ particles can be
accommodated in the state. The $|\nu \rangle _{n}$ is the eigenstate of
number operator $N$:

\begin{equation}
N|\nu \rangle _{n}=\nu |\nu \rangle _{n}.  \label{e4.0}
\end{equation}

Let $a^{\dagger }$ be the creation operator and $b$ the annihilation
operator~(the reason why the creation and the annihilation operators are not
hermitian conjugate will be explained later). Since $n$ is the maximum
occupation number, the creation and annihilation operators$\ a^{\dagger }$
and $b$ satisfy the following conditions:
\begin{equation}
\left\{ \matrix{ a^{\dagger }|n\rangle _{n}=0 & \mbox{  or  } & (a^{\dagger
})^{n+1}|0\rangle _{n}=0, \cr b|0\rangle _{n}=0~~ & \mbox{  or  } &
b^{n+1}|n\rangle _{n}=0.~~~ } \right.  \label{e4.1}
\end{equation}

The relations between creation and annihilation operators for fermions and
bosons are
\begin{equation}
\left\{ \matrix{ aa^{\dagger }-a^{\dagger }a=1 & \mbox{for bosons,}~~ \cr
aa^{\dagger }+a^{\dagger }a=1 & \mbox{for fermions.} } \right.  \label{e4.2}
\end{equation}
The relations that we want to find must return to the cases of bosons and
fermions when $n=\infty $ and $n=1$, so a direct way is to assume the
commutation relation takes the form
\begin{equation}
ba^{\dagger }-f(n)a^{\dagger }b=1,  \label{e4.3}
\end{equation}
and $f(n)$ satisfies
\begin{equation}
f(1)=-1,~~~~~~~~f(\infty )=1.  \label{e4.4}
\end{equation}

For being consistent with these conditions, we choose
\begin{equation}
f(n)=e^{i\frac{2\pi }{n+1}}.  \label{e4.5}
\end{equation}
Thus, the basic commutation relation is
\begin{equation}
ba^{\dagger }-e^{i\frac{2\pi }{n+1}}a^{\dagger }b=1.  \label{e4.5a}
\end{equation}
To coincide with the restrictions of Eqs. (\ref{e4.1}) and (\ref{e4.5a}), the
results of $a^{\dagger }$ and $b$ acting on the state vectors can be chosen
in the following form:
\begin{equation}
\left\{ \matrix{ a^{\dagger }|\nu \rangle _{n}=\sqrt{\frac{1-e^{i2\pi
(\nu +1)/(n+1)}}{1-e^{i2\pi /(n+1)}}}~|\nu +1\rangle
_{n}=\sqrt{\langle \nu +1\rangle _{n}}~|\nu +1\rangle _{n}, \cr b|\nu
\rangle _{n}=\sqrt{\frac{1-e^{i2\pi \nu /(n+1)}}{1-e^{i2\pi
/(n+1)}}}~|\nu -1\rangle _{n}=\sqrt{\langle \nu \rangle _{n}}~|\nu
-1\rangle _{n},~~~~~~~~~ } \right.  \label{e4.6}
\end{equation}
where the following notation is introduced:
\begin{equation}
\langle \nu \rangle _{n}\equiv \frac{1-e^{i2\pi \nu /(n+1)}}{
1-e^{i2\pi /(n+1)}}=\sum_{j=0}^{\nu -1}e^{i2\pi j/(n+1)}.
\label{e4.7}
\end{equation}
According to Eq. (\ref{e4.6}), all the states can be generated from the
ground state $|0\rangle _{n}$ by successive operations of $a^{\dagger }$:
\begin{equation}
|\nu \rangle _{n}=\frac{(a^{\dagger })^{\nu }}{\sqrt{\prod\limits_{j=1}^{\nu
}\langle j\rangle _{n}}}|0\rangle _{n}.  \label{e4.8a}
\end{equation}
It is easy to verify that $_{n}\langle \nu |\nu ^{\prime }\rangle
_{n}=\delta _{\nu \nu ^{\prime }}$.

The commutation relations between number operator and $a^{\dagger }$ and $b$
can be directly derived:
\begin{equation}
\lbrack N,a^{\dagger }]=a^{\dagger },~~~~~~~~~~[N,b]=-b.  \label{e4.8b}
\end{equation}%
Similar to the cases of bosons and fermions, the number operator can be
expressed by the creation and annihilation operators $a^{\dagger }$, $a$, $%
b^{\dagger }$ and $b$. It should be emphasized that the expression of $N$ is
not unique. In fact, in the case of fermion, $b_{f}^{\dagger }b_{f}$ and $%
(b_{f}^{\dagger }b_{f})^{2}$, etc., all can be used as the number operator,
where $b_{f}^{\dagger }$ and $b_{f}$ are the creation and annihilation
operators of fermions. The number operator can be expressed as

\begin{eqnarray}
N &=&\frac{n+1}{2\pi }\arccos \lbrack \frac{1}{2}(ab^{\dagger }+ba^{\dagger
}-a^{\dagger }b-b^{\dagger }a)\rbrack \\
&=&\frac{n+1}{4}-\frac{n+1}{2\pi }\sum\limits_{n=0}^{\infty }\frac{(2n)!}{%
2^{4n+1}(n!)^{2}(2n+1)}(ab^{\dagger }+ba^{\dagger }-a^{\dagger }b-b^{\dagger
}a)^{2n+1}.
\end{eqnarray}

The reason why creation and annihilation operators are not hermitian
conjugate of each other can be realized from Eq. (\ref{e4.6}). Brief
calculation gives
\begin{equation}
a^{\dagger }b|\nu \rangle _{n}=\langle \nu \rangle _{n}|\nu \rangle _{n}.
\label{e4.8c}
\end{equation}
$\langle \nu \rangle _{n}$, the eigenvalue of $a^{\dagger }b$, is not a real
number in general, so $a^{\dagger }b$ is not an hermitian operator. It means
that the creation operator can not be the hermitian conjugate of the
annihilation operator. Only when $n=\infty $ or $n=1$, $\langle \nu \rangle
_{n}$ becomes real, so the operator $a^{\dagger }b$ will be hermitian. In
such cases, we have $(a^{\dagger }b)^{\dagger }=a^{\dagger }b$ and hence $%
a^{\dagger }=b^{\dagger }$. This is just the case of bosons or fermions.

For considering the conjugate operators of $a^{\dagger }$ and $b$, taking
the hermitian conjugate of Eq. (\ref{e4.5a}), we have
\begin{equation}
ab^{\dagger }-e^{-i\frac{2\pi }{n+1}}b^{\dagger }a=1.  \label{e4.8d}
\end{equation}
Furthermore, a set of relations for $a$ and $b^{\dagger}$ analogous to Eq. (
\ref{e4.6}) can be obtained easily replacing $\langle \nu \rangle
_{n}$ and $\langle \nu+1 \rangle _{n}$ by their complex conjugate. The
results show that $b^{\dagger }$ and $a$ can play roles similar to
creation and annihilation operators $a^{\dagger }$ and $b$.

According to the results of the operators acting on the state vectors, the
following operator relations among $a^{\dagger }$, $b$, $b^{\dagger }$ and $%
a $ can be obtained:
\begin{equation}
\matrix{ a=b^{\ast }, \cr a^{\dagger }a=b^{\dagger
}b,~~~~~aa^{\dagger }=bb^{\dagger }, \cr aaa^{\dagger }+a^{\dagger
}aa=2\cos {\frac{\pi }{n+1}}~aa^{\dagger }a,~ \cr bbb^{\dagger
}+b^{\dagger }bb=2\cos {\frac{\pi }{n+1}}~bb^{\dagger
}b,\mbox{etc.} }  \label{e4.9}
\end{equation}
and
\begin{equation}
\matrix{ a^{\dagger }a|\nu \rangle _{n}=b^{\dagger }b|\nu \rangle _{n}=
|\langle \nu \rangle _{n}||\nu \rangle _{n}~~~~~~ \cr aa^{\dagger }|\nu
\rangle _{n}=bb^{\dagger }|\nu \rangle _{n}= |\langle \nu+1 \rangle
_{n}||\nu \rangle _{n}. }
\end{equation}

\section*{3. Representation of algebra of angular momentum}

So far we have obtained an operator realization of Gentile statistics in
which one quantum state can be occupied by $n$ particles. Based on this
result, we now consider the realization of angular momenta.

Straightforwardly, one can check that for the case of $n=1$ (the fermion
case) we have

\begin{equation}
\left\{ \matrix{ J_{+}=a^{\dagger},~~~~ \cr J_{-}=a,~~~~~ \cr
J_{z}=N-\frac{n}{2}, } \right.
\end{equation}
and

\begin{eqnarray}
\lbrack J_{+},J_{-}] &=&2J_{z},  \label{e5.3} \\
\lbrack J_{z},J_{\pm }] &=&\pm J_{\pm }.~  \nonumber
\end{eqnarray}%
Obviously, the relations among the operators $J_{+}$, $J_{-}$ and $J_{z}$
are the same as those of angular momentum ${\bf J}=J_{x}{\bf i}+J_{y}{\bf j}
+J_{z}{\bf k}$ (where $J_{\pm }=J_{x}\pm iJ_{y}$). It should be emphasized
that, this result only holds for $n=1$. In the case of $n=1$, the maximum
occupation number is $1$ and hence there are only two possible states: $%
|0\rangle _{n=1}$ and $|1\rangle _{n=1}$. This means that corresponding to
the case $n=1$, the magnitude of the angular momentum $j$ must be $\frac{1}{2
}$ since for $j=\frac{1}{2}$ there are also only two states: $|+\frac{1}{2}
\rangle $ and $|-\frac{1}{2}\rangle $.

Similarly, we can also construct a realization for angular momenta $j=1$ in
terms of the creation and annihilation operators corresponding to $n=2$
since they both have three states: $|+1\rangle $, $|0\rangle $, $|-1\rangle $
and $|0\rangle _{n=2}$, $|1\rangle _{n=2}$, $|2\rangle _{n=2}$. It can be
verified in a straightforward fashion that in this case
\begin{equation}
\left\{ \matrix{ J_{+}=\sqrt{2}a^{\dagger }, \cr J_{-}=\sqrt{2}a,~ \cr
J_{z}=N-\frac{n}{2}, }\right.
\end{equation}%
and $J_{\pm }$, $J_{z}$ satisfy the relations Eq. (\ref{e5.3}).

These results imply an underlying relation between the magnitude of the
angular momentum $j$ and the maximum occupation number $n$: To realize the
angular momentum ${\bf J}$, one needs to use the creation and annihilation
operators which correspond to $n=2j$. Then, we wish, therefore, to construct
realizations for the other angular momenta.

For $j=\frac{3}{2}$, we can chose

\begin{equation}
\left\{
\matrix{
J_{+}=\lambda _{1}^{\ast }a^{\dagger}+\lambda _{2}^{\ast}b^{\dagger}, \cr
J_{-}=\lambda _{1}a+\lambda _{2}b,~~ \cr
J_{z}=N-\frac{n}{2},~~~~~~~ } \right.
\end{equation}
where

\begin{eqnarray}
\lambda _{1} &=&\frac{1}{2\sqrt{2}}(2+2^{3/4})(1+i), \nonumber \\
\lambda _{2} &=&\frac{1}{2^{1/4}}-1.
\end{eqnarray}
These relations hold for $n=3$.

For $j=2$ and $n=4$:

\begin{equation}
\left\{
\matrix{
J_{+}=\lambda _{1}^{\ast }a^{\dagger}+\lambda _{2}^{\ast}b^{\dagger}, \cr
J_{-}=\lambda _{1}a+\lambda _{2}b,~~ \cr
J_{z}=N-\frac{n}{2},~~~~~~~ } \right.
\end{equation}
where

\begin{eqnarray}
\lambda _{1} &=&-\frac{\sqrt{2}}{4}\sqrt{5\sqrt{5}-5+\sqrt{182\sqrt{5}-370}}-i%
\frac{\sqrt{2}}{4}\sqrt{5+3\sqrt{5}+\sqrt{62+\frac{158}{\sqrt{5}}}},\nonumber \\
\lambda _{2} &=&\sqrt{\sqrt{5}-\sqrt{\frac{22}{\sqrt{5}}-5} }.
\end{eqnarray}
In the above cases, the operators $J_{+}$ and $J_{-}$ can be expressed as a
linear combination of the creation and annihilation operators. However, for $%
j\geq \frac{5}{2}$ and $n\geq 5$ , we need to use high order terms of the
creation and annihilation operators such as $a^{\dagger}aa^{\dagger}$. For
instance, in the case of $j=\frac{5}{2}$ and $n=5$

\begin{equation}
\left\{
\matrix{
J_{+}=\lambda _{1}^{\ast }a^{\dagger}+\lambda _{2}^{\ast}b^{\dagger}+
\lambda _{3}^{\ast }a^{\dagger}aa^{\dagger}, \cr
J_{-}=\lambda_{1}a+\lambda _{2}b+\lambda _{3}aa^{\dagger}a,~~~ \cr
J_{3}=N-\frac{n}{2},~~~~~~~~~~~~~~~~~~~~~ } \right.
\end{equation}
where

\begin{eqnarray}
\lambda _{1} &=&\sqrt{\frac{9}{2}-\frac{1}{\sqrt{3}}}-i3^{1/4}, \nonumber \\
\lambda _{2} &=&0, \\
\lambda _{3} &=&i\sqrt{\frac{1}{\sqrt{3}}-\frac{1}{2}}.\nonumber
\end{eqnarray}
Also, for other values of $j$ and $n$, we can obtain the realizations for
other angular momentum operators.

In terms of creation and annihilation operators corresponding to Gentile
statistics, we have constructed a kind of realization for angular momentum ($%
SU(2)$) algebra. It is important to notice that such a realization scheme is
not unique.

\section*{4. Discussion}

At the beginning of this paper, we have compared two kinds of
representations (the Holstein-Primakoff and the Schwinger representations)
of the algebra of angular momentum, and pointed out that if one wants to
realize the algebra of angular momentum by means of only one set of creation
and annihilation operators, he must necessarily make a departure from
Bose-Einstein statistics and make use of a kind of intermediate statistics
just like what has been done in the Holstein-Primakoff representation. One aim
of this paper is to provide an operator realization of algebra of angular
momentum. To achieve this, we begin by first showing that we can provide an
operator realization of Gentile statistics (a kind of immediate statistics
in which one quantum state can be occupied by $n$ particles). Based on such
operator relations for the creation and annihilation operators, we construct
a representation of the algebra of angular momentum ($SU(2)$ algebra).

Different $n$ (the maximum occupation number) corresponds to different kinds
of statistics. The two special cases are $n=1$ and $\infty $, which are
Bose-Einstein statistics and Fermi-Dirac statistics. Moreover, for the other
values of $n$, e.g. $n=2,3...$, we have various kinds of intermediate
statistics. In Section 3 we find that the angular momentum with magnitude $j=%
\frac{1}{2}$ can be expressed by the creation and annihilation operators
corresponding to $n=1$, and moreover, the case $j=1$ corresponds to $n=2$,
the case $j=\frac{3}{2}$ corresponds to $n=3$ and so on. The parameter $n$
is physically meaningful, it labels various kinds of statistics. The
result implies that there exist some underlying connections between the
angular momentum ($SU(2)$) algebra and the operator realization of Gentile
statistics.

The representation for angular momentum introduced above is based on the
operator realization of Gentile statistics. Clearly, the particles which
obey Gentile statistics must not be real particles. In other words, this
representation, just as the case in the Holstein-Primakoff representation, is
related to a kind of imaginary particles. It can be expected that, this
result can be used to deal with some complex interaction systems.

\vskip 0.5cm

We would like to thank Dr. Yong Liu for sending us some important
references, and we are very indebted to Dr. G. Zeitrauman for his
encouragement. This work is supported in part by LiuHui fund.

\end{document}